\documentclass[epj]{svjour}
\usepackage{amsmath}
\usepackage{amssymb}
\usepackage{bm}
\usepackage{graphicx}
\begin{document}
\title{Lensing effects in a nematic liquid crystal with topological defects}
\author{Caio S\'atiro
\and  Fernando Moraes}
\institute{Departamento de F\'{\i}sica,
           Universidade Federal da Para\'{\i}ba\\
           Caixa Postal 5008, 58051-970,
            Jo\~ao Pessoa, PB, Brazil\\ }

\abstract{
Light traveling through a  liquid crystal with disclinations perceives a geometrical background which causes lensing effects similar to the ones predicted for cosmic objects like global monopoles and cosmic strings. In this article we explore the effective geometry as perceived by light in such media. The comparison between both systems suggests that experiments can be done in the laboratory to simulate optical properties, like gravitational lensing, of cosmic objects.
}
\maketitle
  
\section{Introduction}
Topological defects may be formed when a symmetry is broken in a phase transition. Examples run from Cosmology to Condensed Matter Physics. In Cosmology, defects like global monopoles and cosmic strings might have appeared at the beginning of the universe when the symmetries of the unified forces of nature were broken in a sequence of phase transitions \cite{brand} while the universe cooled down. In Condensed Matter Physics, the prototype topological defect is the Abrikosov-Nielsen-Olesen vortex \cite{abri} in type II superconductors with broken U(1) symmetry. On the other hand, uniaxial nematic liquid crystals, in their isotropic phase, have global SO(3) rotational symmetry, which is broken under cooling, yielding the nematic phase and producing  topological defects like the hedgehog point defect or disclinations as relics of the former, less ordered, phase. Remarkably, they are formed according to the same physical principle, the Kibble mechanism \cite{kibble}, as the cosmic defects, as verified experimentally by Bowick {\it et al.} \cite{bowick}. 

The similarity between defects in nematics and cosmic defects goes beyond their processes of formation. Cosmic strings, for instance, although do not introduce curvature (gravitational field) globally, they do change the topology and, therefore, introduce interesting phenomena like gravitational lensing \cite{uzam}. In this article, we use the geometrical model  developed by Joets and Ribotta \cite{joets} for  anisotropic media to study  the propagation of light in the nematic phase with topological defects, finding lensing effects. This result, in fact, is not new. It has been previously observed as early as 1919 by Grandjean \cite{grand} in a different context. Our aim in this work is to reveal the  geometric structure underlying this system and its similarity to cosmological defects, which permits that experiments may be performed with liquid crystals to study otherwise unreachable objects like the global monopole or the cosmic string, in the spirit of reference \cite{bowick}.

Another system that has been compared with the cosmic string is the irrotational hydrodynamical vortex \cite{visser}. Propagation of phonons in a medium with such vortices can be described by an effective Riemannian geometry which asymptotically agrees with that of the massless spinning cosmic string. The phonon rays are geodesics in this effective geometry and a converging lens behavior has been found \cite{visser}. 

\section{Geometry}
In an inhomogeneous, isotropic, medium Fermat's principle, stating that the optical length of a ray between two given points be the shortest one joining these points,  is nicely interpreted as a geodesic in a Riemannian space \cite{livro}. A generalization of Riemannian geometry, Riemann-Cartan geometry, has been a valuable tool to the study of topological defects both in Gravitation and in Condensed Matter Physics \cite{bjp}. Yet another generalization of Riemannian geometry, Finsler geometry \cite{finsler}, has been used to model the liquid crystal as an optical medium \cite{joets}. This geometry takes into account anisotropy by considering that the line element depends not only on the location but also on the local orientation of the line to be considered. This seems a very reasonable approach to light propagation through anisotropic media since the local orientation of the molecules is incorporated by the geometry perceived by the light rays. Accordingly, Joets and Ribotta \cite{joets} interpreted the light rays in an anisotropic medium as  geodesics in a Finsler space. Although this ``anisotropic'' Riemannian geometry appear as a natural, powerful, tool to model such problem, ordinary Riemannian geometry is enough \cite{caio} for highly symmetric molecular configurations like the ones presented by topological defects in nematics. For the cases studied here  Riemannian geometry gives exactly the same answers as the more general Finslerian approach. Therefore, for simplicity,  we present our calculations  in the Riemannian approach in the body of this article, leaving the Finslerian approach for the Appendix.

It is important here to make a distinction between ordinary and extraordinary light rays propagating in uniaxial nematics:  in the former the electric field is perpendicular to the director $\vec{n}$ and in the latter the electric field has a component along $\vec{n}$. Differently from the ordinary ray, in an extraordinary ray, the direction of the Poynting vector $\vec{S}$ in general differs from the direction of the wavevector $\vec{k}$. This implies that the energy velocity differs from the phase velocity, leading to two different refractive indices \cite{livro}: the ray index $N_r$, associated to the energy velocity, and the phase index $N_p$, associated to the phase velocity, given by
\begin{equation}
N_r^2=n_o^2\cos^2\beta +n_e^2\sin^2\beta , \label{nr}
\end{equation}
where $\beta = (\widehat{\vec{n},\vec{S}})$ and
\begin{equation}
N_p^2=\frac{n_o^2 n_e^2}{n_e^2\cos^2\gamma +n_o^2\sin^2\gamma},
\end{equation}
where $\gamma = (\widehat{\vec{n},\vec{k}})$. In both equations $n_o$ and $n_e$ are, respectively, the ordinary and extraordinary indices \cite{livro}.

Among all possible paths between the generic points $A$ and $B$, Fermat's principle for the extraordinary rays grants us that the path actually followed by the energy is the one that minimizes the integral (see, for example, section 3.3.3.2 of \cite{kleman})
\begin{equation}
{\cal F}=\int_{A}^{B} N_{r} d \ell . \label{fermat}
\end{equation}
Here, $d\ell$ is the element of arc length along the path. Ordinary rays are not interesting because their refractive index is simply $n_o$, a constant.

In Riemannian geometry the line element $ds$ depends on the position coordinates $x^i$ of the point of the manifold under consideration
\begin{equation} 
ds^2 = \sum_{i,j}g_{ij}dx^idx^j. \label{riemline}
\end{equation}
The geodesic joinning points $A$ and $B$ in such manifold is obtained by minimizing $\int ds$, just like Fermat's principle. This leads to a nice interpretation of the light paths as geodesics in an effective geometry \cite{livro}.   Thus, we may identify
\begin{equation} 
N_{r}^{2}d\ell^2  = \sum_{i,j}g_{ij}dx^idx^j. \label{interp}
\end{equation}
The meaning of this equation is the following: the line element of the optical path in Euclidean space with refractive properties is identified with the line element of an effective geometry characterized by $g_{ij}$. 

The equation for the rays is the geodesic equation in Riemannian space \cite{manfredo},
\begin{equation}
\frac{d^{2}x^i}{dt^2}+\sum_{j,k}\Gamma^{i}_{jk}\frac{dx^j}{dt}\frac{dx^k}{dt}=0,\label{georie}
\end{equation}
where $t$ is a parameter along the geodesic and $\Gamma^{i}_{jk}$ are the Riemannian connections, or Christoffel symbols, given by
\begin{equation}
\Gamma^{i}_{jk}=\frac{1}{2}\sum_{m}g^{mi}\left\{ \frac{\partial g_{km}}{\partial x^j}+\frac{\partial g_{mj}}{\partial x^k}-\frac{\partial g_{jk}}{\partial x^m}\right\} . \label{chris}
\end{equation}

\begin{figure}[b]
\hspace{0.5in}
\includegraphics[height=6cm]{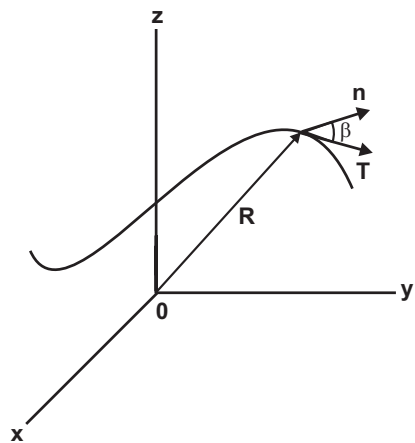} 
\label{linha}
\caption{The curve $\vec{R}(\ell)$ in three-dimensional space, its tangent vector $\vec{T}(\ell)$ and the director $\vec{n}$.}
\end{figure}

Among the many geometrical properties that derive from the metric, the Riemann tensor and the curvature scalar are the most important because they give a feeling for how curvature is distributed in space. For the specific expressions for the Riemann tensor and the curvature scalar in terms of the metric and Christoffel symbols, we refer the reader to any good textbook on Riemannian geometry such as \cite{manfredo}.

The effective metric $g_{ij}$ felt by the extraordinary light propagating in the medium with ray index $N_r$ is obtained from Eq. (\ref{interp}). The angle $\beta$ is determined from the specific configurations of the director field $\vec{n}$ as follows. If the curve $\vec{R}(\ell)$, where $\ell$ is a parameter along the curve, represents the light trajectory, then
\begin{equation}
\vec{T}(\ell)=\frac{d\vec{R}}{d\ell} \label{tangent}
\end{equation}
is the tangent vector to the curve at each position parametrized by $\ell$ (see Fig. 1). It is a well known result of the differential geometry of curves (see for example section II B of \cite{kamien}) that $\|\vec{T}(\ell)\|=1$ if $\ell$ is the arc length, which is the case. Since $\|\vec{n}\|$=1, we have that 
\begin{equation}
\cos \beta = \vec{T}\cdot \vec{n}. \label{cosbeta}
\end{equation}

\begin{figure}[!b]
\begin{center}
\includegraphics[height=1.2in]{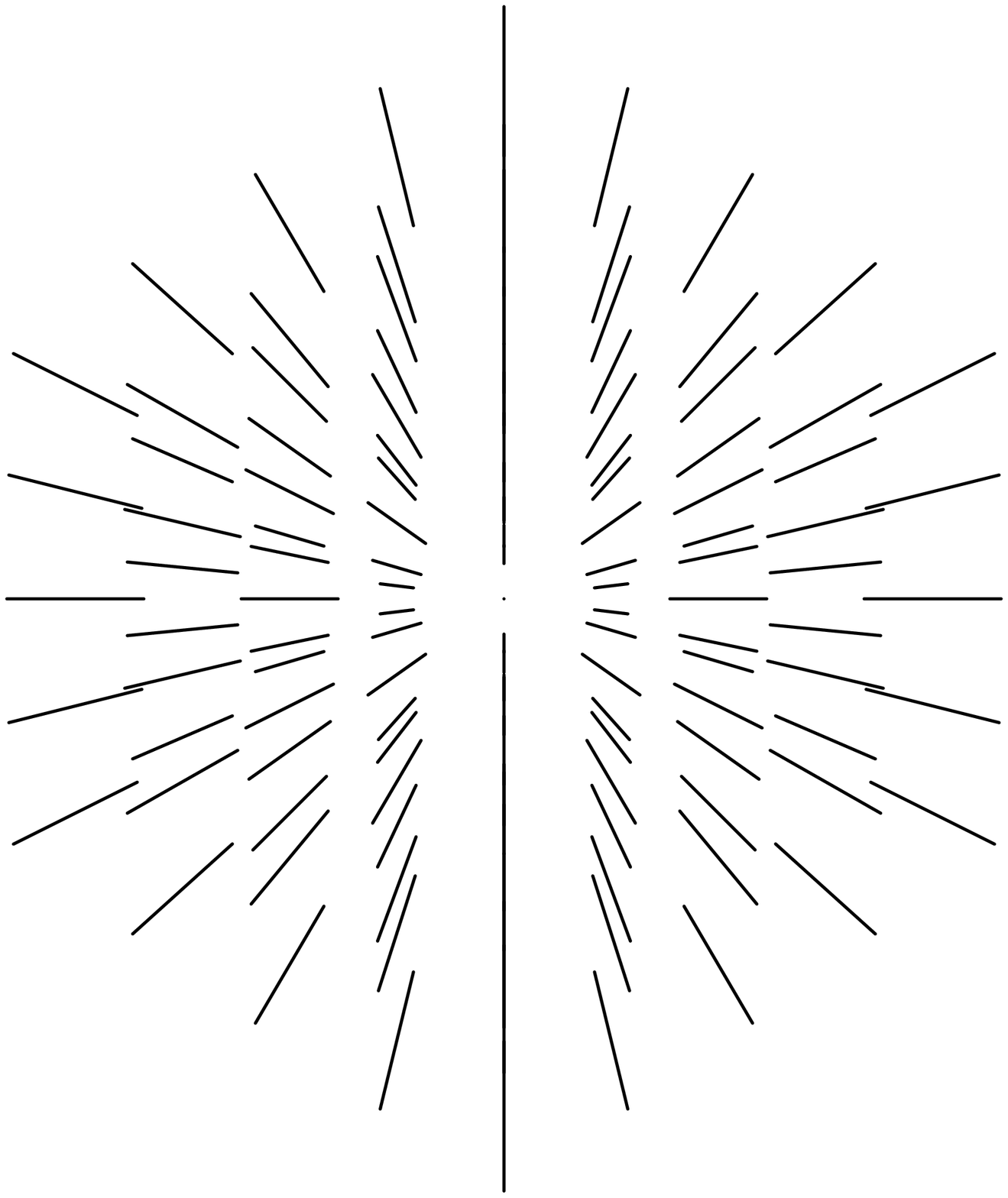} 
\includegraphics[height=1.2in]{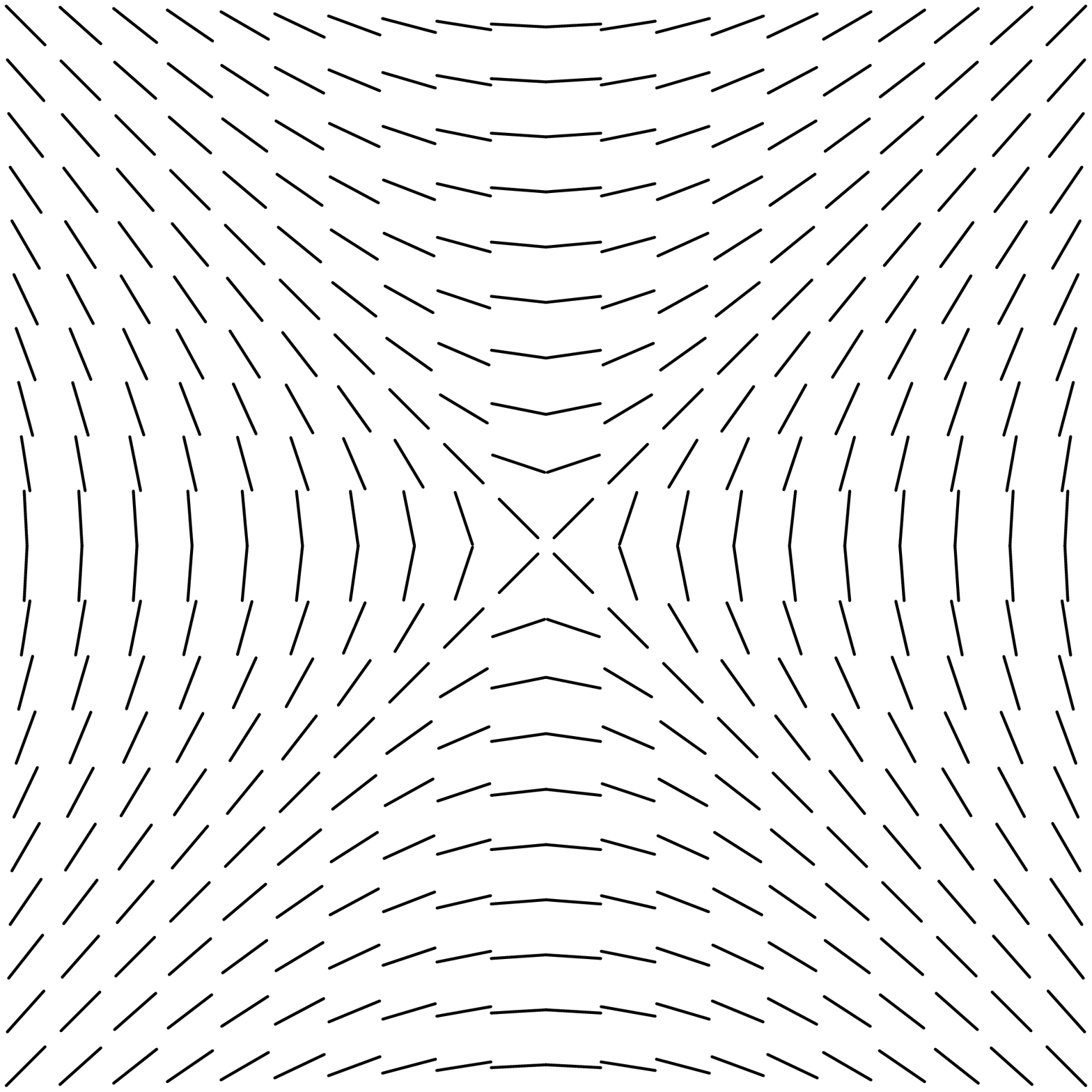}
\caption{Director field for the hedgehog point defect and the $k=-1$, $c=\frac{\pi}{2}$ disclination, respectively.}
\includegraphics[height=1.2in]{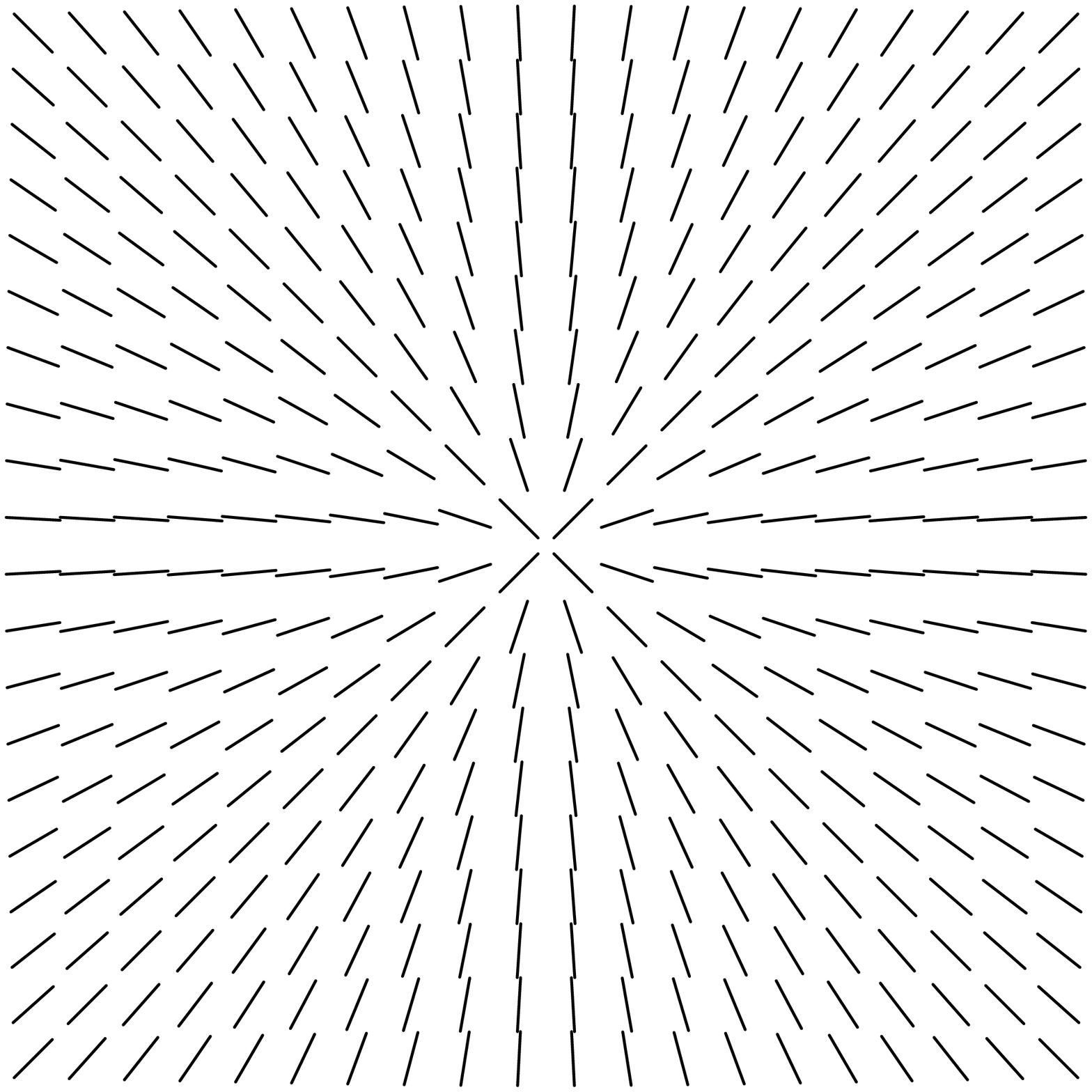} 
\includegraphics[height=1.2in]{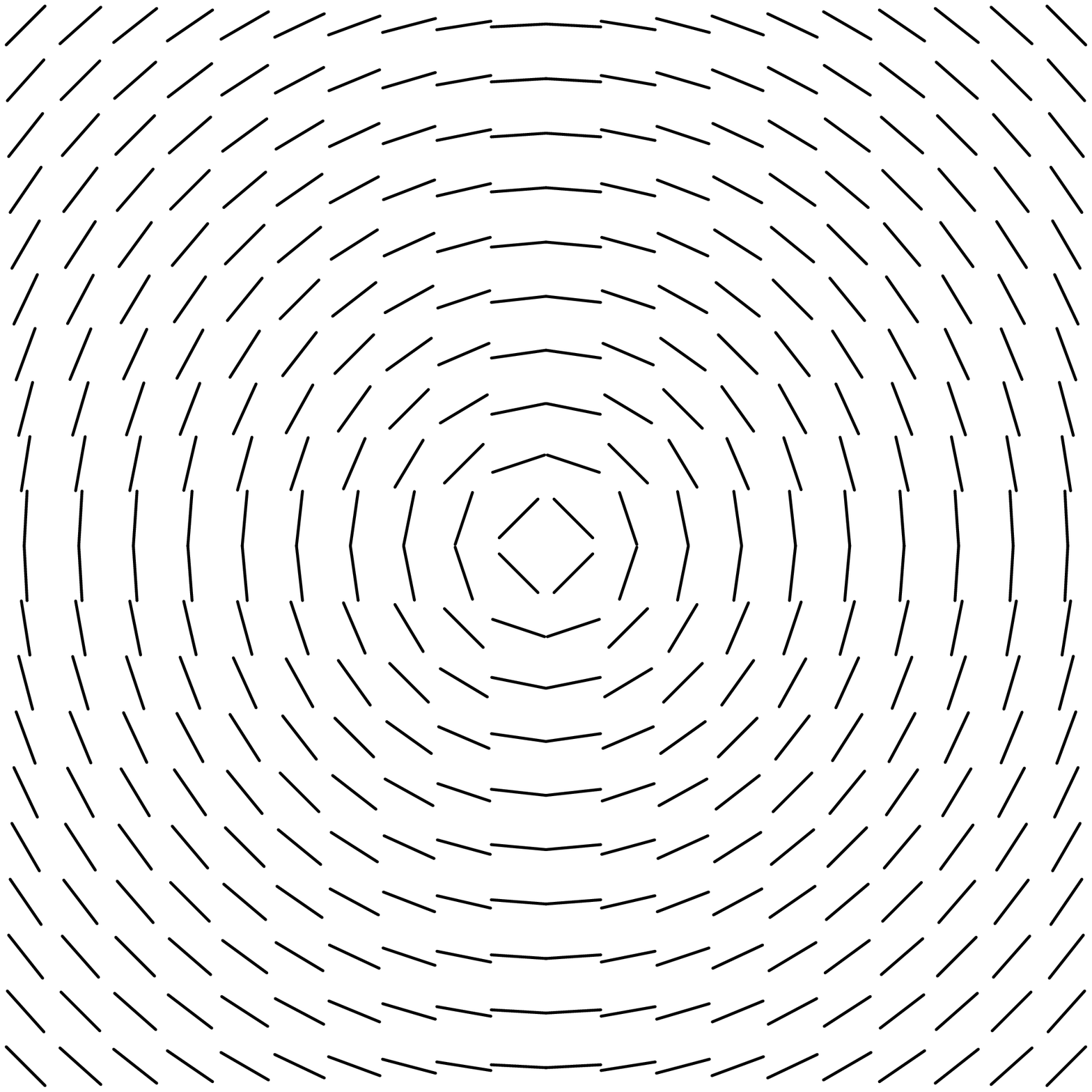} 
\caption{Director field  for the strength $k=1$, $c=0$ and $k=1$, $c=\frac{\pi}{2}$ disclinations, respectively.}
\includegraphics[height=1.2in]{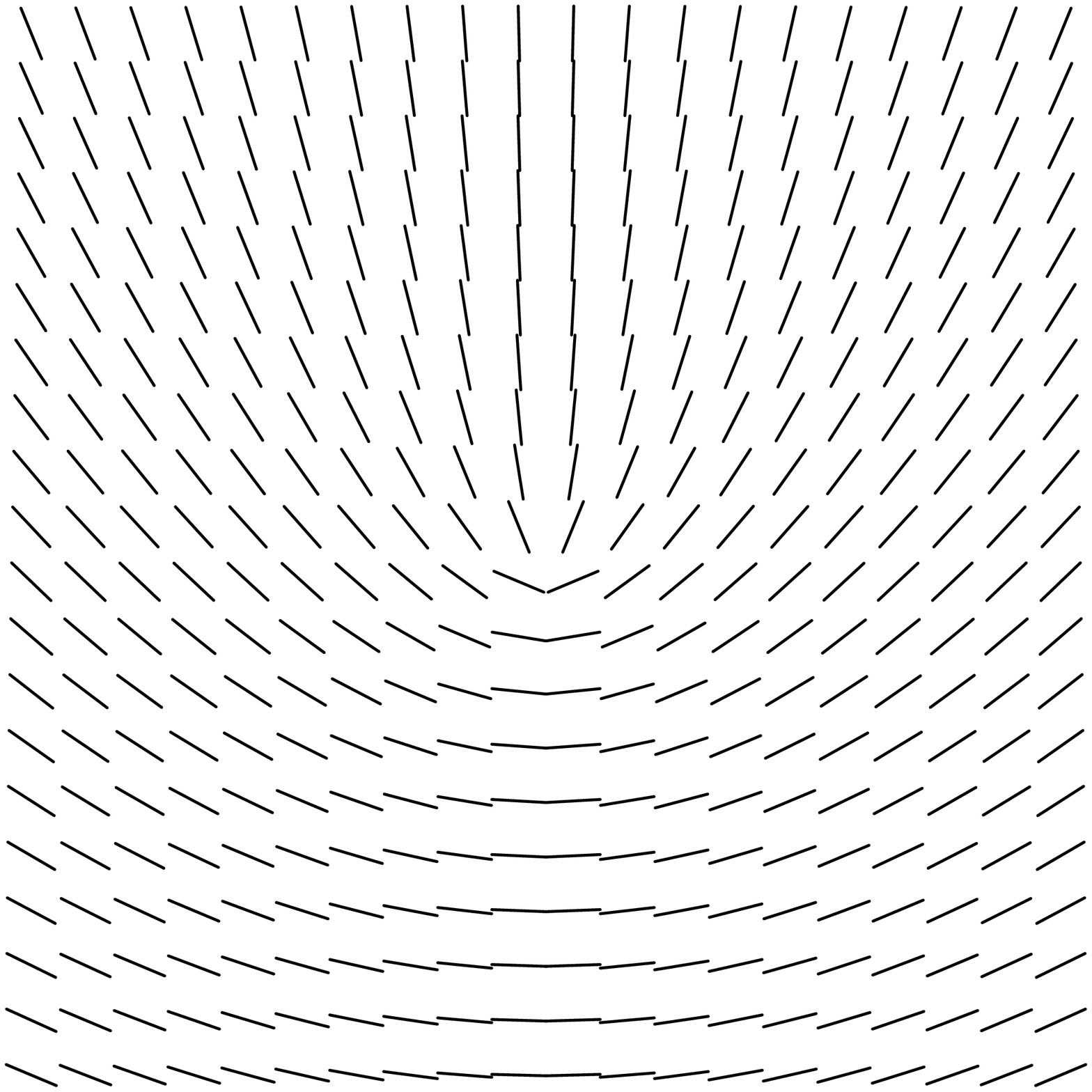} 
\includegraphics[height=1.2in]{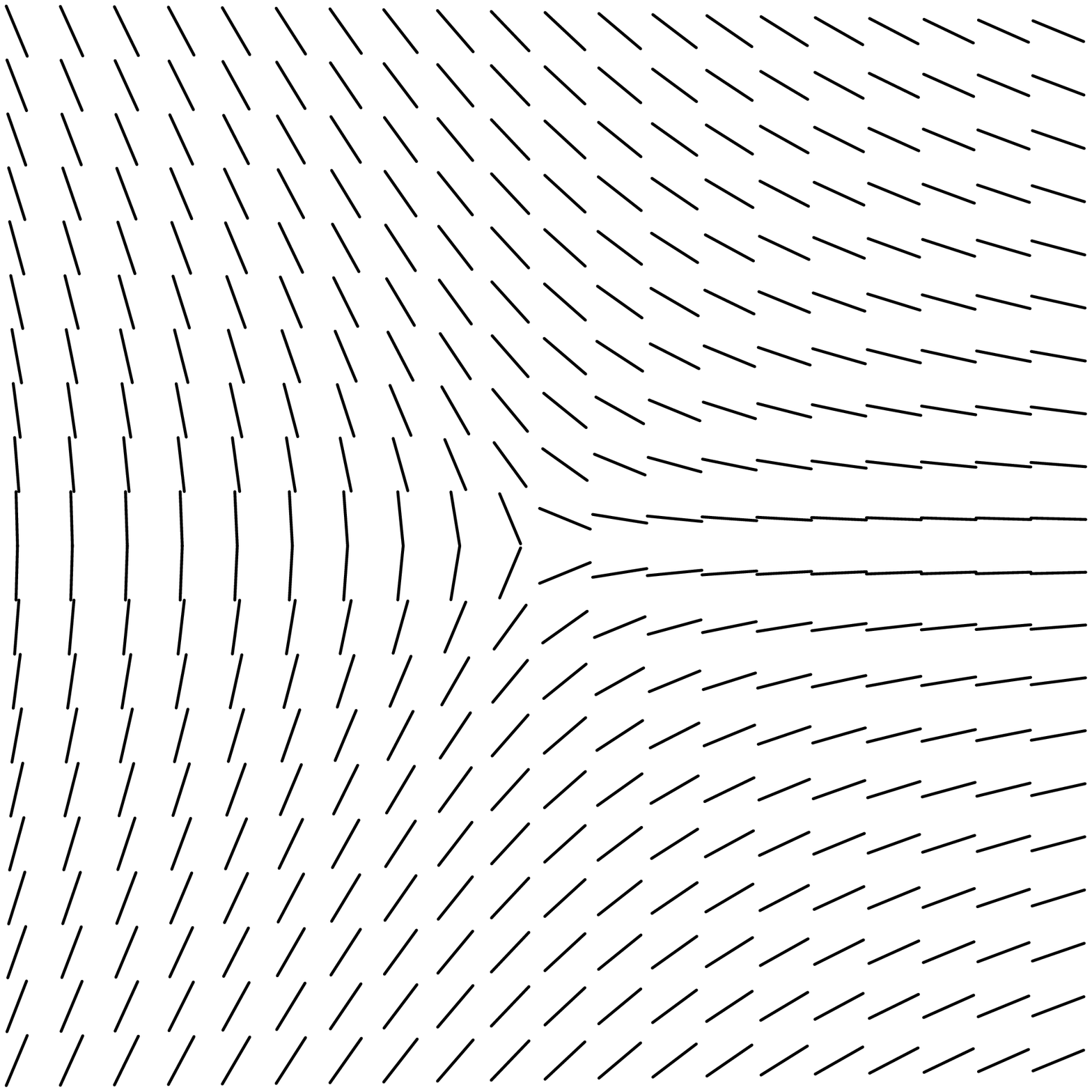} 
\caption{Director field for the $k=\frac{1}{2}$, $c=\frac{\pi}{4}$ and $k=-\frac{1}{2}$, $c=0$ disclinations, respectively.}
\label{campos}
\end{center}
\end{figure}

We study the hedgehog point defect and wedge disclinations of generic strength $k$ and constant c. The hedgehog director configuration is characterized by the unitary radial vector
\begin{equation}
\vec{n}=\hat{r} \label{point}
\end{equation}
in spherical coordinates, whereas the disclination configurations in the plane $z=const$, are given by 
\begin{equation}
\vec{n}=(\cos\varphi,\sin\varphi,0), \label{nline}
\end{equation}
in Cartesian coordinates,
where \cite{kleman}
\begin{equation}
\varphi=k\phi+c , \label{phi}
\end{equation}
where $\phi$ is the angular polar coordinate. Selected director configurations can be seen in Figs. 2 - 4.

It is assumed that the optical axis lies along the average molecular orientation.  Also, assuming the line defects straight, their translational symmetry along their axis allows us to ignore the dimension along the defect and study a two-dimensional cross section of the medium, the plane $z=const$. A word of caution here: the $k=\pm 1$ line defects are known to be unstable with respect to a escape into the third dimension \cite{kleman}. Therefore, they might be very difficult to be observed. In fact, recent results \cite{chiccoli} indicate that even in thin films the escaped texture is the stable configuration. One might think that director fields similar to the $k=\pm 1$ defects could be  obtained with the help of electric or magnetic fields or even by surface anchoring  schemes but, to the best of our knowledge, this has not been observed so far. So, the results concerning the  $k=\pm 1$ defects are, in this respect, just a theoretical possibility. For experimental procedures the hedgehog and $k=\pm \frac{1}{2}$ line defects will be more interesting due to their known stability \cite{kleman}.

\subsection{Hedgehog}
Since the hedgehog is spherically symmetric we use spherical coordinates and write $\vec{R}=r\hat{r}$, in the spherical coordinate basis $\{\hat{r},\hat{\theta},\hat{\phi}\}$. Thus,
\begin{equation}
\vec{T}=\frac{d\vec{R}}{d\ell}=\frac{dr}{d\ell}\hat{r}+r\frac{d\hat{r}}{d\ell}.
\end{equation}
But, using  the Euclidean line element in spherical coordinates,
\begin{equation}
d\ell^2=dr^2+r^{2}(d\theta^{2}+\sin^{2}\theta d\phi^{2}) \label{eucl}
\end{equation}
and Eq. (\ref{point}) and, since $\frac{d\hat{r}}{d\ell}$ is orthogonal \cite{kamien} to $\hat{r}$, Eq. (\ref{cosbeta}) becomes
\begin{equation}
\cos\beta=\dot{r}, \label{cosb}
\end{equation}
leading to
\begin{equation}
\sin\beta=r\sqrt{\dot{\theta}^2+\sin^2\theta\dot{\phi}^2}, \label{sinb}
\end{equation}
where $\dot{r}=\frac{dr}{d\ell}$, $\dot{\theta}=\frac{d\theta}{d\ell}$ and $\dot{\phi}=\frac{d\phi}{d\ell}$.

Eqs. (\ref{cosb}), (\ref{sinb}), with the help of Eq. (\ref{nr}), make Eqs. (\ref{riemline}) and (\ref{interp}) into
\begin{equation}
ds^2=N_{r}^{2}d\ell^2  = n_{o}^2 dr^2 + n_{e}^2 r^{2}(d\theta^{2}+\sin^{2}\theta d\phi^{2}).\label{hedgemetr}
\end{equation}

Now, if we rescale the radial coordinate by making $\rho=n_{o}r$ the metric (\ref{hedgemetr}) becomes
\begin{equation}
ds^2 =d\rho^{2} + \alpha^2 \rho^{2}(d\theta^{2}+\sin^{2}\theta d\phi^{2}),  \label{monometr}
\end{equation}
with $\alpha = n_e/n_o$. Notice that, in the isotropic limit, $n_o=n_e$, the Euclidean line element in spherical coordinates, Eq. (\ref{eucl})  is recovered. The metric (\ref{monometr}) is identical to the space section of the global monopole metric \cite{mono}.

The nonzero Riemann tensor components corresponding to the metric (\ref{hedgemetr}) is given by
\begin{eqnarray}
R_{\theta\phi\theta\phi}= -\frac{n_{e}^2(n_{e}^2-n_{o}^2)}{n_{o}^2}r^2\sin ^2 \theta
\end{eqnarray}
and the permutations $R_{\phi\theta\phi\theta}=-R_{\phi\theta\theta\phi}=-R_{\theta\phi\phi\theta}=R_{\theta\phi\theta\phi}$.
The curvature scalar is
\begin{equation}
R=\frac{2(n_{e}^2-n_{o}^2)}{n_{e}^2 n_{o}^2 r^2}.
\end{equation}
Again, in the $n_o=n_e$ limit, the Euclidean space (null curvature everywhere) is recovered.

In summary, the  effective geometry perceived by light in the presence of a hedgehog point defect, as described by its metric, Riemann tensor and curvature scalar, coincides with the space part of the geometry associated to a global monopole \cite{mono}, a topological defect of spacetime.

\subsection{Disclinations}
Since we consider light propagating in the transversal $z=const$. surface to the line defects, we write $\vec{R}=r\cos\phi\hat{x}+r\sin\phi\hat{y}$, combining the polar coordinates $r$ and $\phi$ with the Cartesian basis $\{\hat{x},\hat{y}\}$. Following the same procedure as in the previous subsection, we get 
\begin{equation}
\vec{T}=(\dot{r}\cos\phi -r\dot{\phi}\sin\phi )\hat{x}+(\dot{r}\sin\phi+r\dot{\phi}\cos\phi )\hat{y}, \label{tline}
\end{equation}
where, again, $\dot{r}=\frac{dr}{d\ell}$ and $\dot{\phi}=\frac{d\phi}{d\ell}$. Now, with the help of Eqs. (\ref{nline}) and (\ref{tline}), Eq. (\ref{cosbeta}) becomes
\begin{equation}
\cos\beta=\dot{r}\cos (\varphi -\phi )+r\dot{\phi}\sin (\varphi -\phi ), \label{cosline}
\end{equation}
which implies that
\begin{equation}
\sin\beta=\dot{r}\sin (\varphi -\phi )+r\dot{\phi}\cos (\varphi -\phi ). \label{sinline}
\end{equation}

Now, using  Eqs. (\ref{riemline}),(\ref{interp}), (\ref{nr}), (\ref{cosline}) and (\ref{sinline}) and the two-dimensional Euclidean line element in polar coordinates
\begin{equation}
d\ell^2=dr^2+r^{2} d\phi^{2} \label{pol}
\end{equation}
we get
\begin{eqnarray}
& ds^2 & = \left\{n_o^2 \cos^{2}[(k-1)\phi+c]+n_e^2 \sin^{2}[(k-1)\phi+c]\right\}dr^{2} \nonumber\\
& + & \left\{n_o^2 \sin^{2}[(k-1)\phi+c]+n_e^2 \cos^{2}[(k-1)\phi+c]\right\}r^{2}d\phi^{2} \nonumber\\
& - & \left\{2(n_e^2-n_o^2)^{2}\sin[(k-1)\phi+c]\cos[(k-1)\phi+c]\right\}rdrd\phi . \nonumber\\
& & \label{kmetric}
\end{eqnarray}

For the metric (\ref{kmetric}) the only nonzero Riemann curvature tensor components are given by
\begin{eqnarray}
R_{r\phi r\phi}= -k(k-1)(n_e^2-n_o^2)\cos \{2[(k-1)\phi+c]\} . \label{tensor}
\end{eqnarray}
and permutations $R_{\phi r\phi r}=-R_{\phi r r\phi}=-R_{r\phi \phi r}=R_{r\phi r\phi}$.
The curvature scalar is
\begin{equation}
R=\frac{2k(k-1)(n_e^{2}-n_o^{2})}{n_o^{2}n_e^{2}r^{2}}\cos \{2[(k-1)\phi+c]\}. \label{scalar}
\end{equation}

By inspection of Eqs. (\ref{kmetric}), (\ref{tensor}) and (\ref{scalar}) we see that, in the isotropic limit, $n_o=n_e$, as in the hedgehog case, we recover Euclidean space. Also, for $n_o \neq n_e$ $k=1$ and $c=0$ and $c=\frac{\pi}{2}$ we reproduce the result of \cite{caio}, which associates the effective geometry of the $k=1$ defects with that of a cosmic string.  A similar result was found for disclinations in elastic solids in the geometric theory of defects approach \cite{kata}. In particular, the $k=1$ and $c=0$ case coincides also with the $\theta=\pi/2$ section of the hedgehog effective metric (Eq. (\ref{hedgemetr})
\begin{equation}
ds^2 =n_{o}^2 dr^{2} + n_{e}^2 r^{2}d\phi^{2}.
\end{equation}
This implies that light propagating in the presence of the hedgehog, on the surface $\theta=\pi/2$, has the same behavior as light propagating on the surface $z=const$ of the $k=1$, $c=0$ disclination

\begin{figure}[!b]
\begin{center}
\includegraphics[height=1.2in]{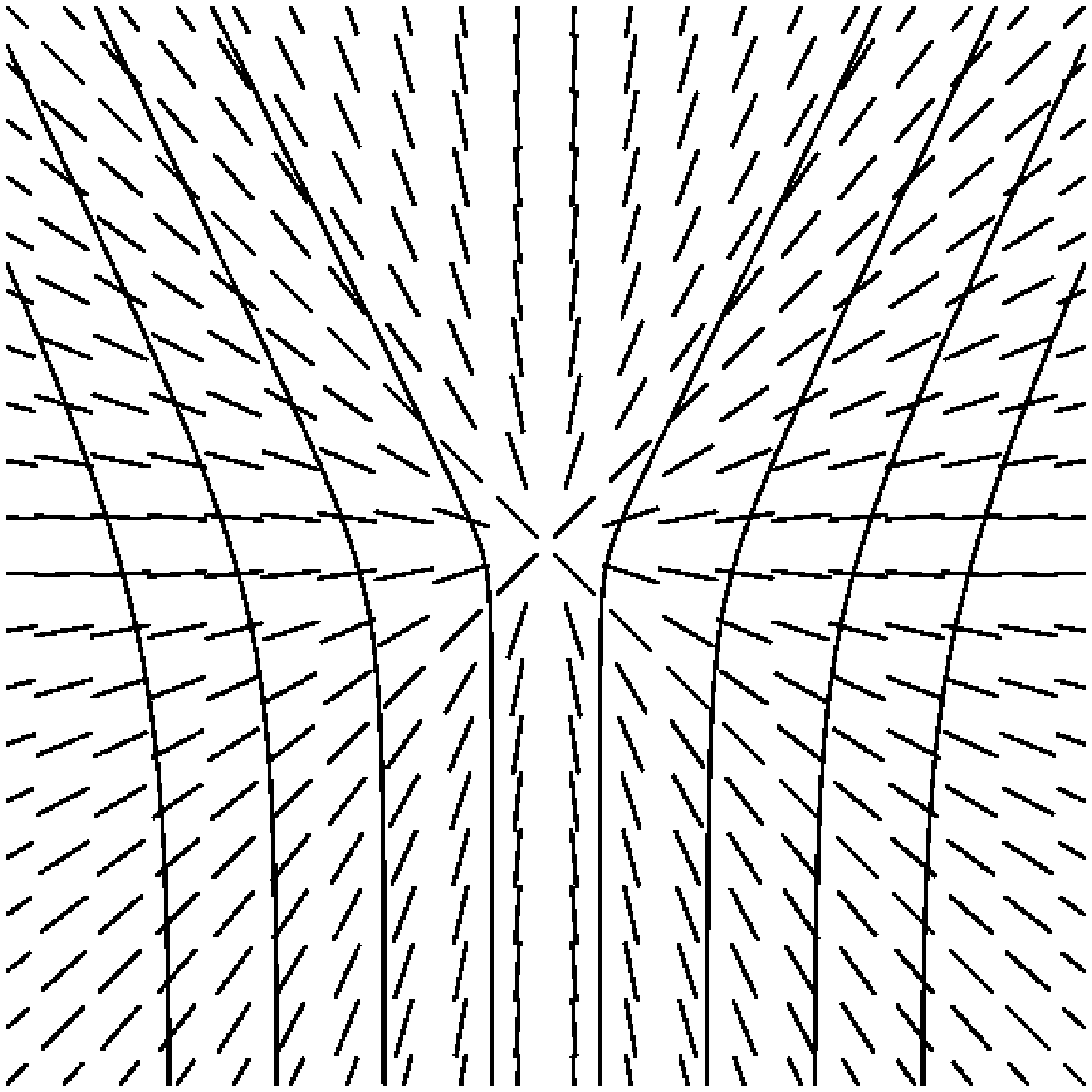} 
\includegraphics[height=1.2in]{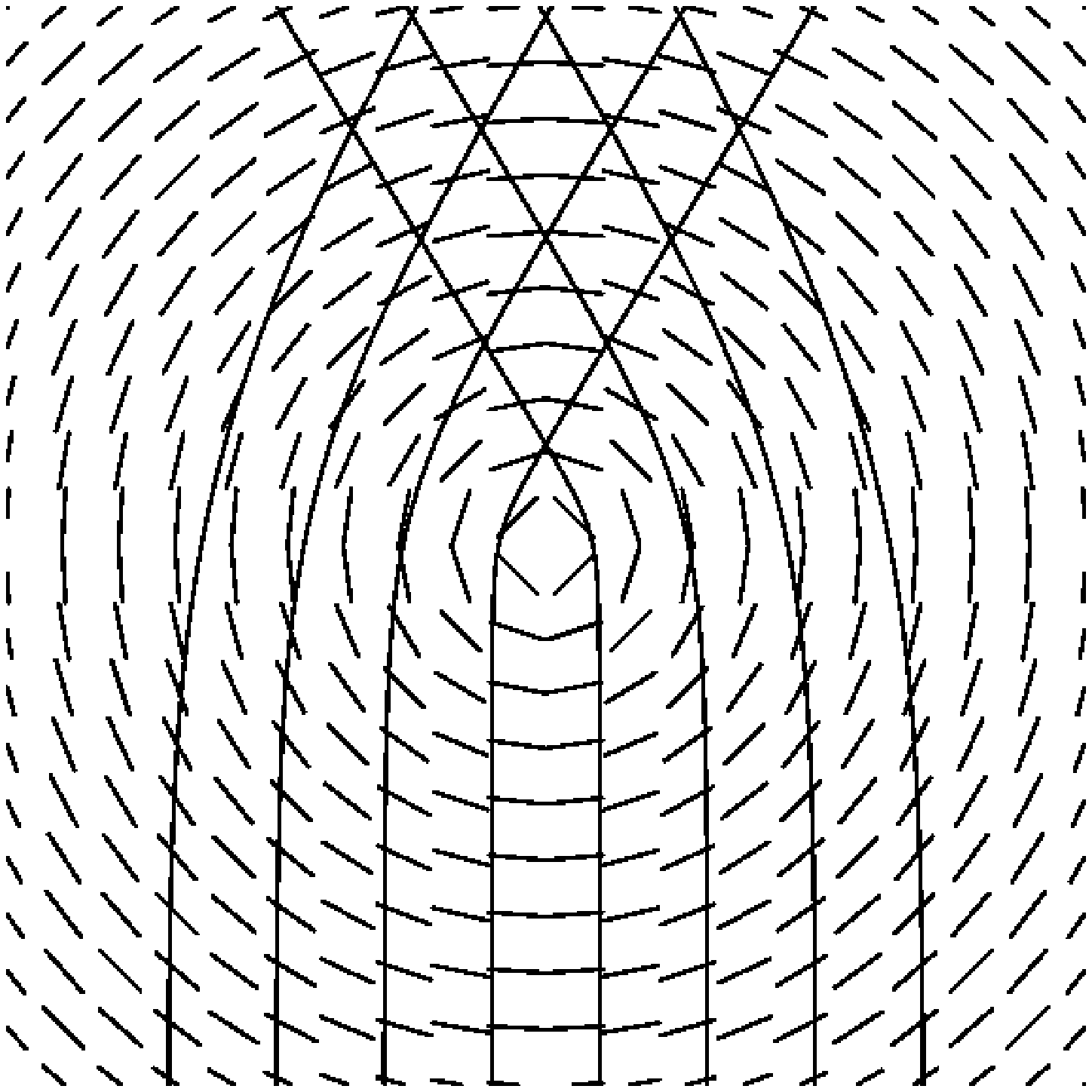} \label{kum}
\caption{Director field  and light paths for the $\theta=\pi/2$ section of the hedgehog (same as  $k=1$, $c=0$ disclination) and $k=1$, $c=\frac{\pi}{2}$ disclination, respectively.}
\includegraphics[height=1.2in]{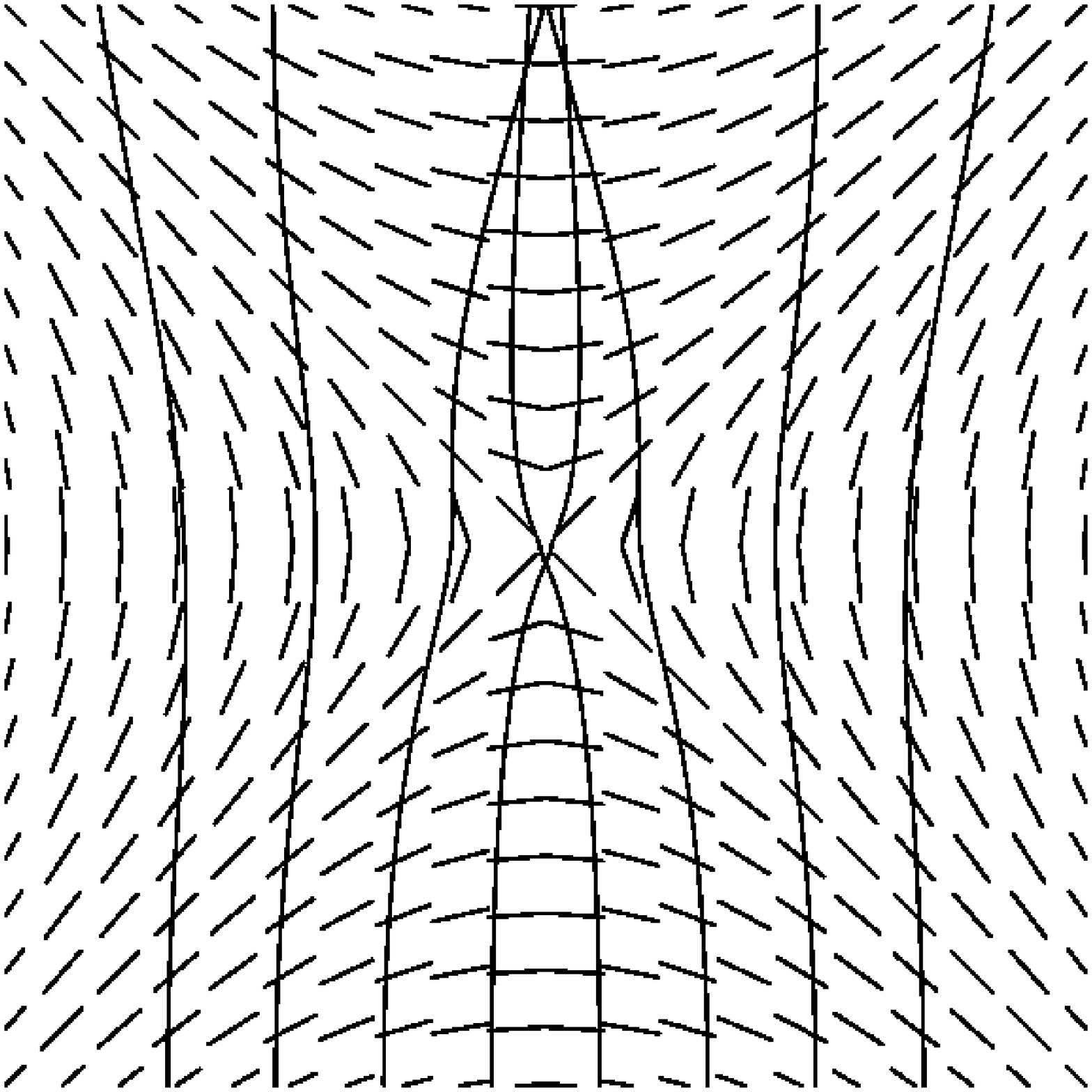} 
\includegraphics[height=1.2in]{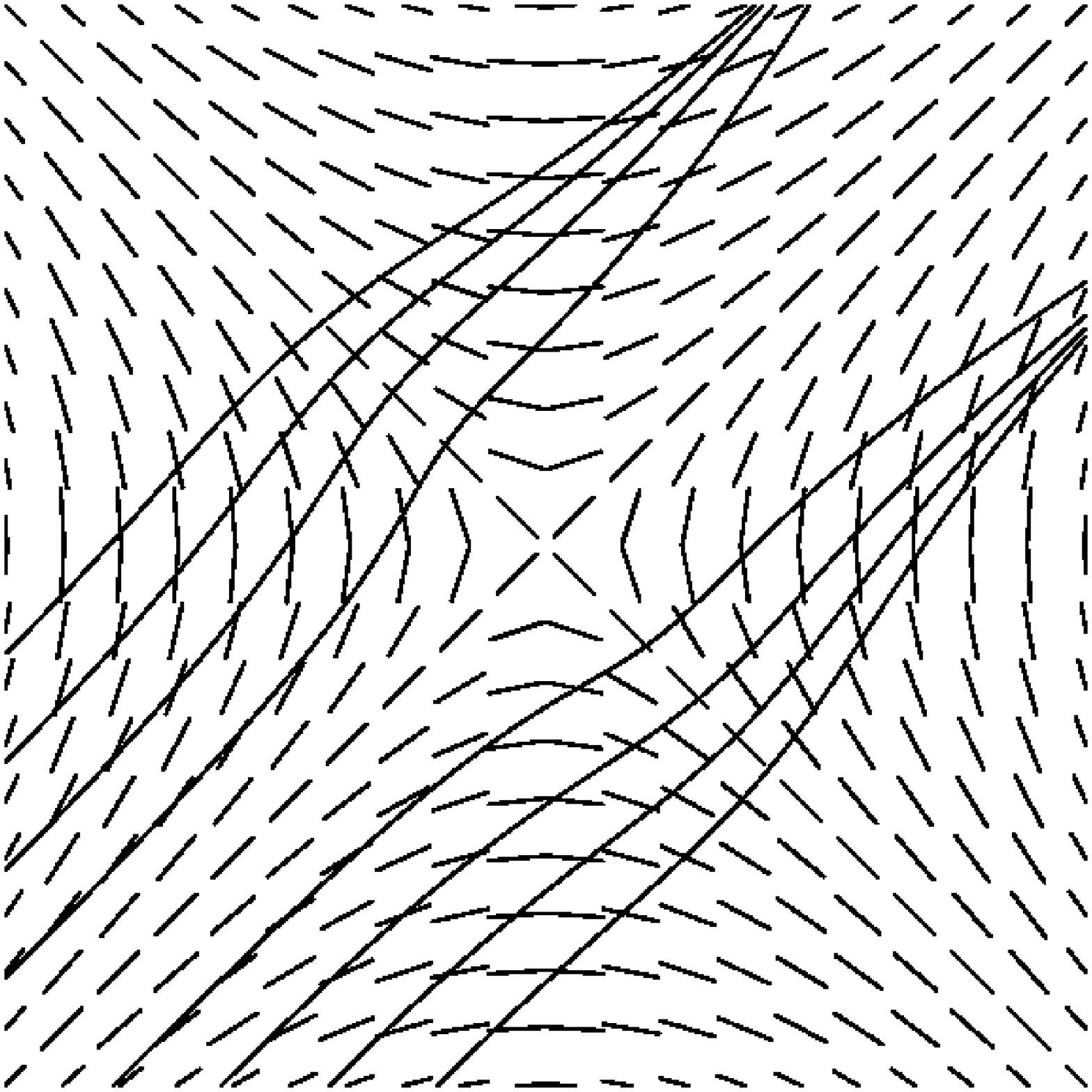}  \label{kmenosum}
\caption{Director field and light paths for the $k=-1$, $c=\frac{\pi}{2}$ disclination.}
\includegraphics[height=1.2in]{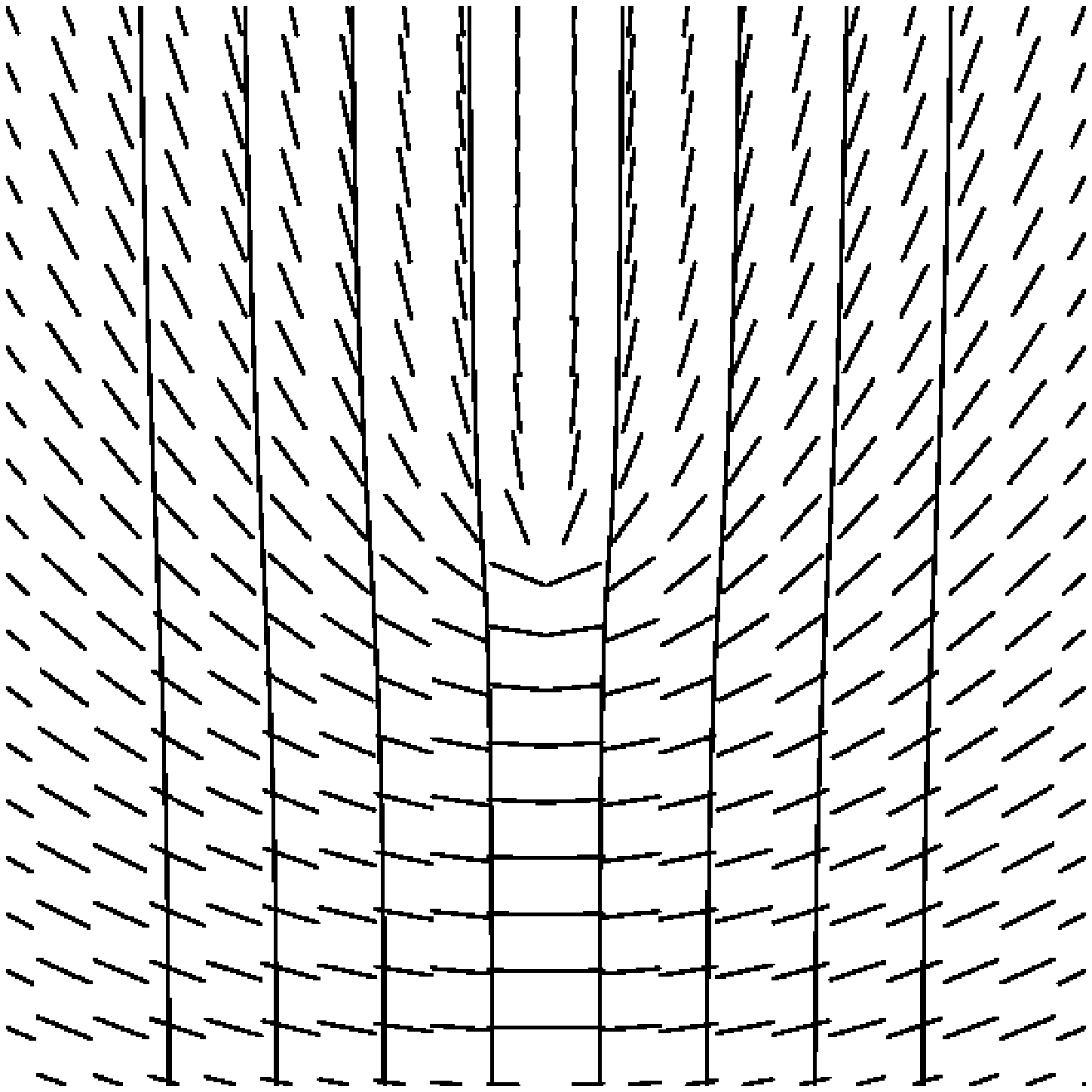} 
\includegraphics[height=1.2in]{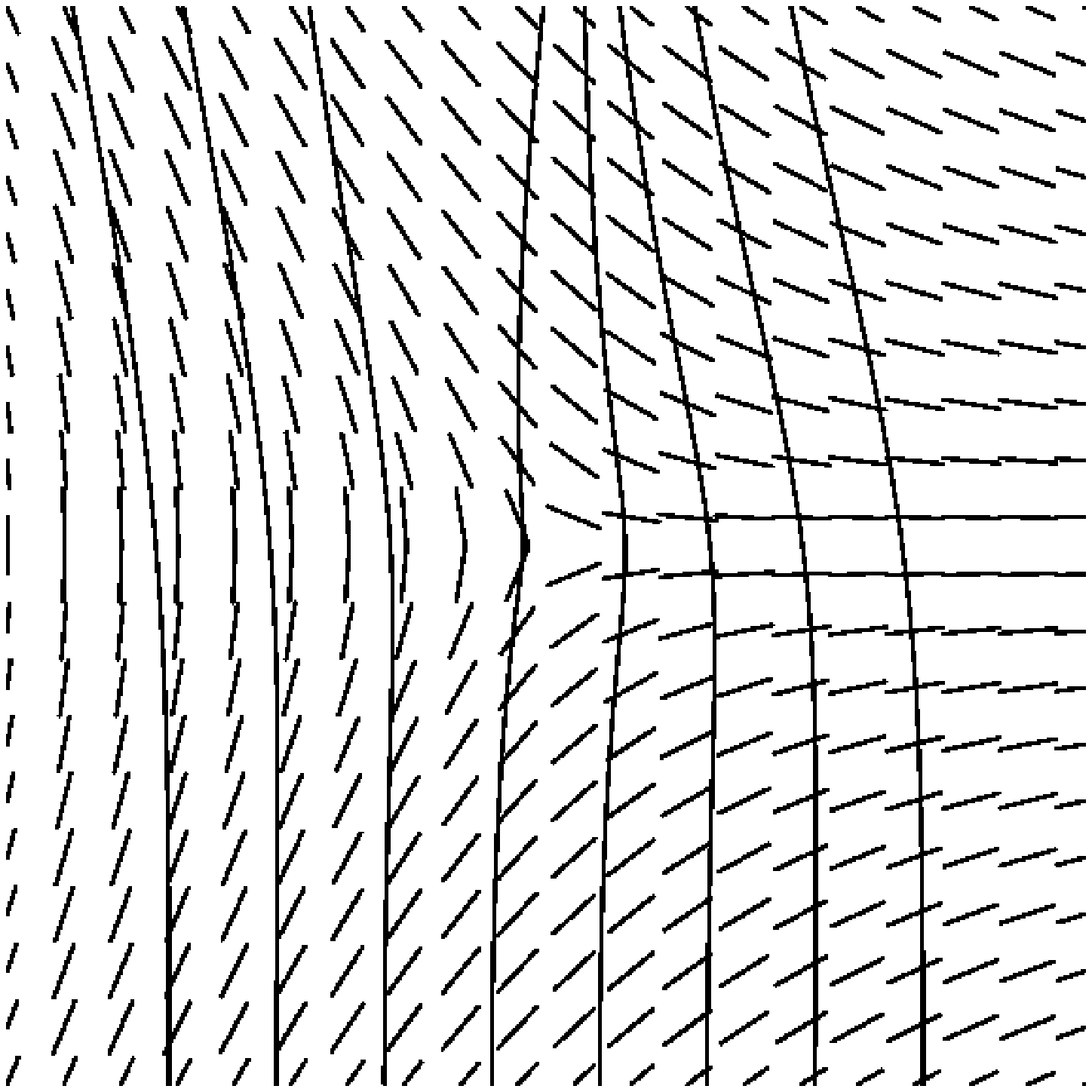}  \label{kmeiovert}
\caption{Director field and light paths (vertical) for $k=\frac{1}{2}$, $c=\frac{\pi}{4}$ and $k=-\frac{1}{2}$, $c=0$ disclinations, respectively.}
\includegraphics[height=1.2in]{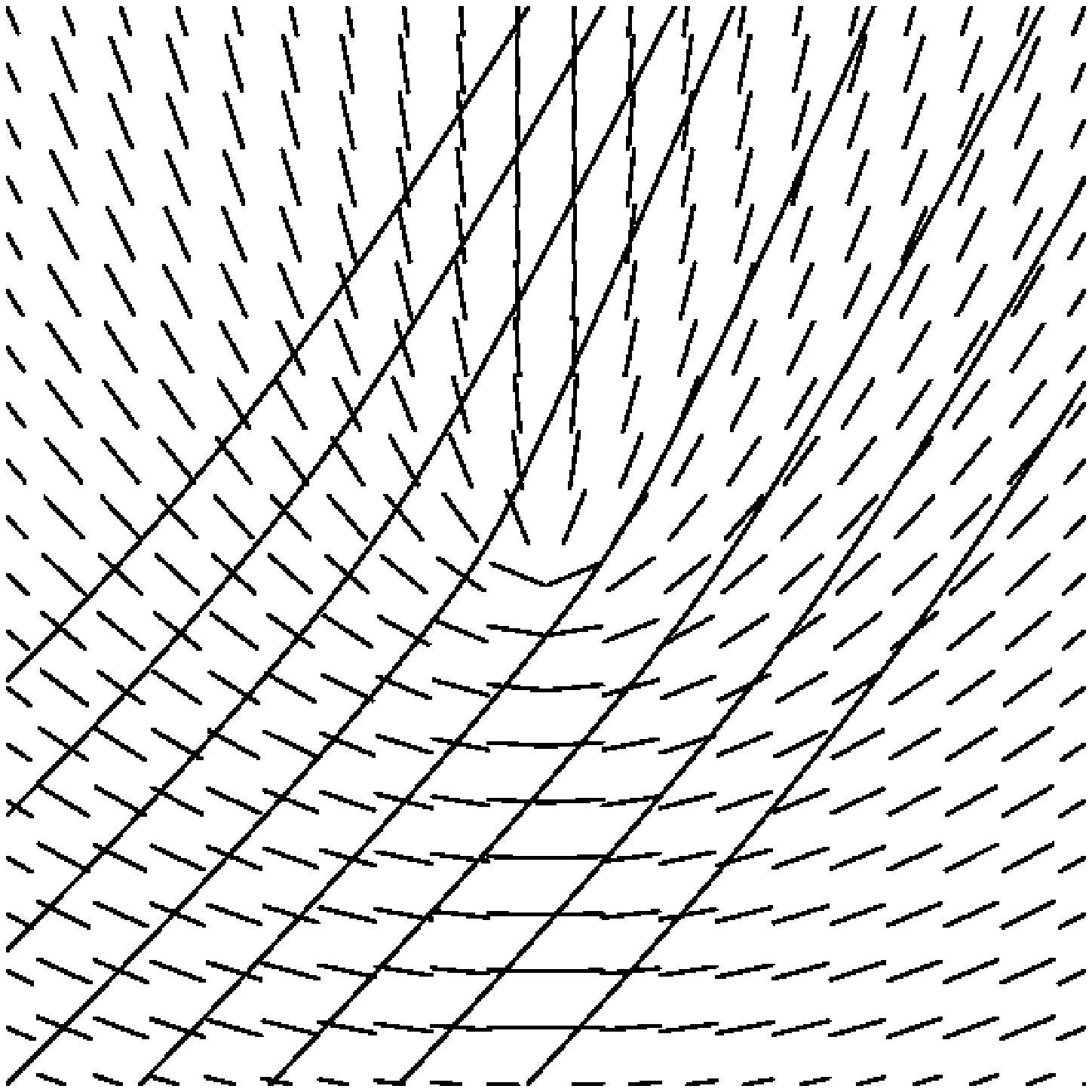} 
\includegraphics[height=1.2in]{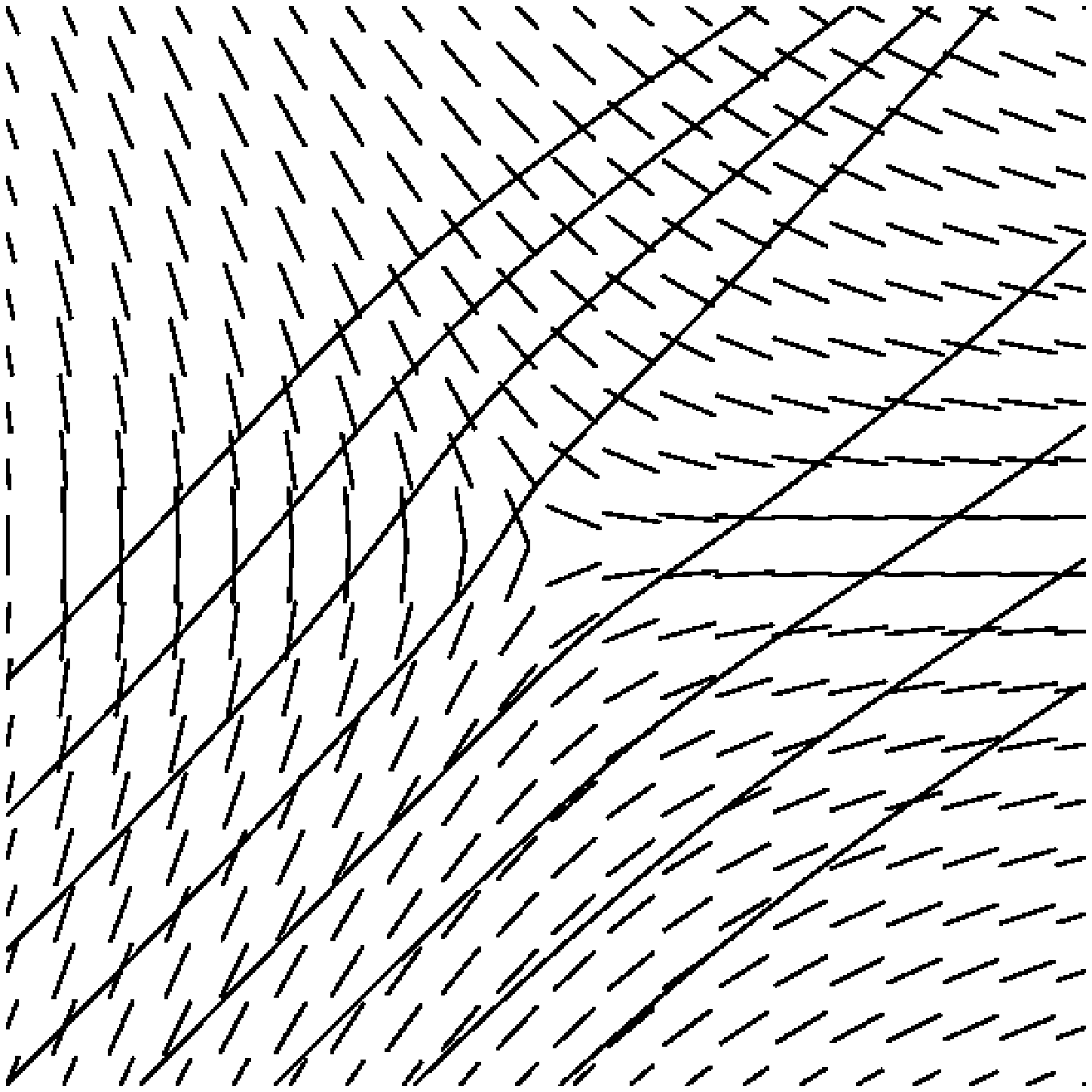}  \label{kmeioskew}
\caption{Director field and light paths (skew) $k=\frac{1}{2}$, $c=\frac{\pi}{4}$ and $k=-\frac{1}{2}$, $c=0$ disclinations, respectively.}
\includegraphics[height=1.2in]{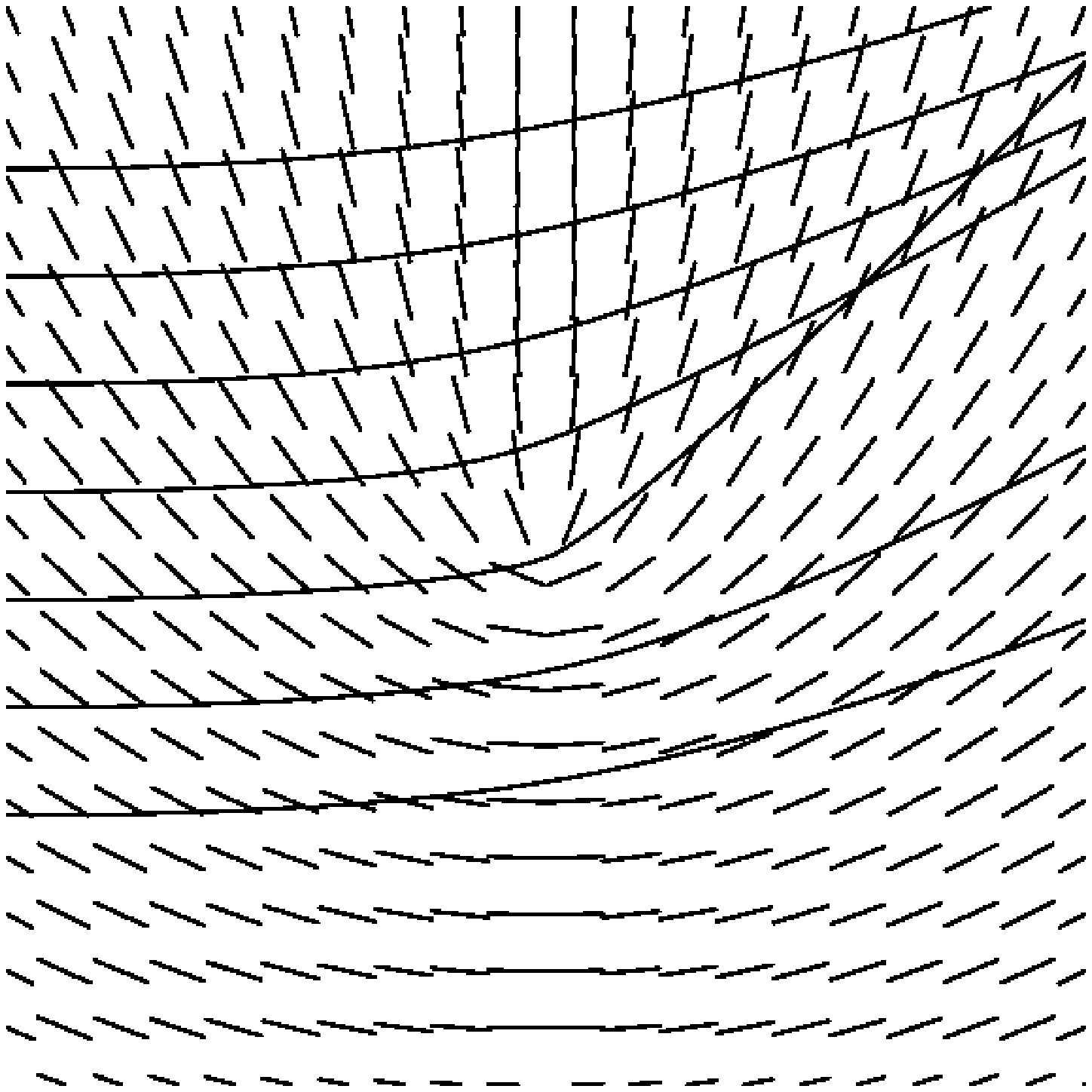} 
\includegraphics[height=1.2in]{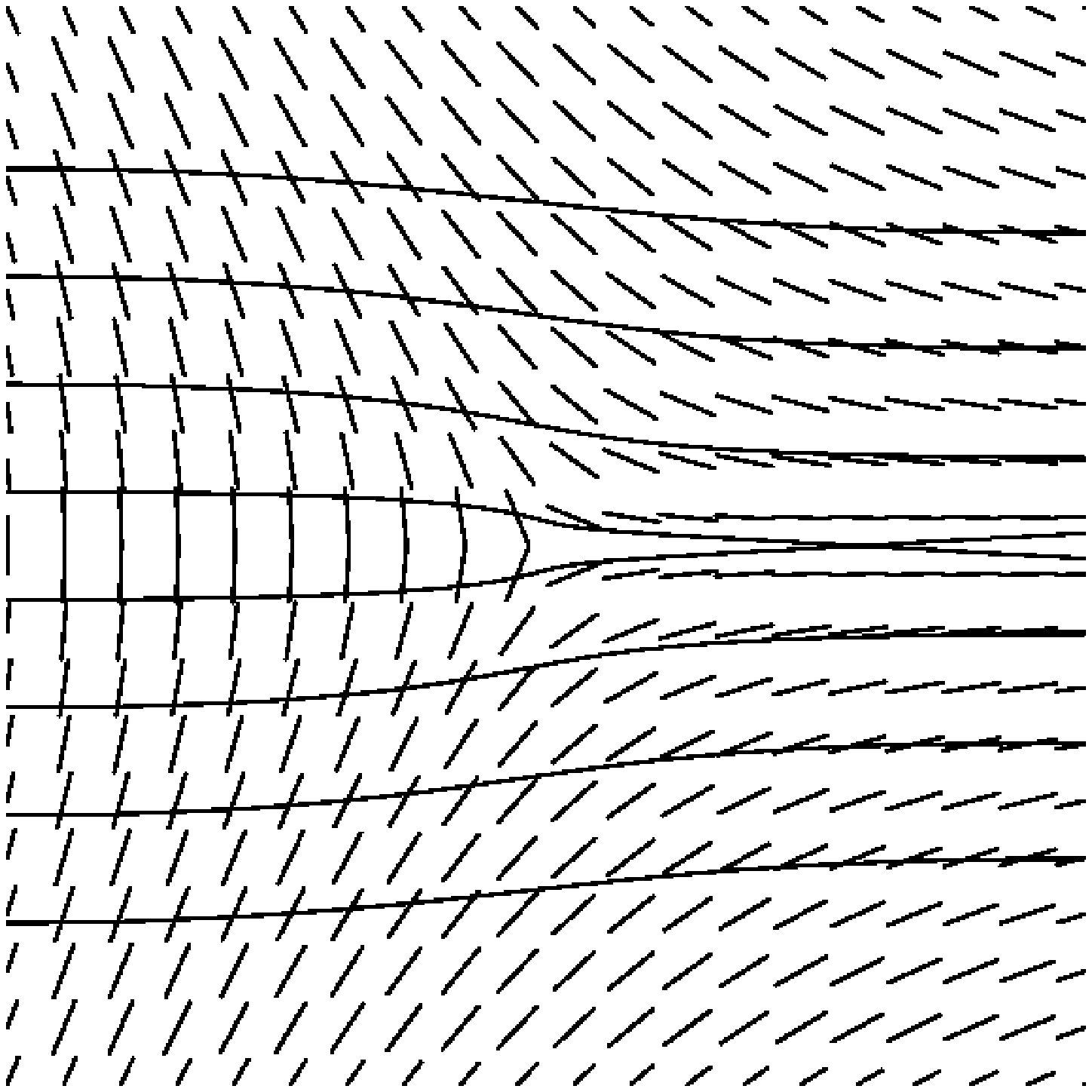} \label{kmeiohoriz}
\caption{Director field and light paths (horizontal) $k=\frac{1}{2}$, $c=\frac{\pi}{4}$ and $k=-\frac{1}{2}$, $c=0$ disclinations, respectively.}
\end{center}
\end{figure}

\section{Light Paths}
The paths followed by light traversing the director configurations corresponding to the defects studied are given by the solutions of Eq. (\ref{georie}) These equations were obtained in their final form after the computation of the Christoffel symbols (Eq. (\ref{chris})) obtained from the metrics (Eqs. (\ref{hedgemetr}) and (\ref{kmetric})  and solved numerically using the Runge-Kutta method.   For the numerical calculation we used the same values as reference \cite{joets}: $n_o=1.65$ and $n_e=1.94$ corresponding to the nematic Merck Phase V and look only at the extraordinary wave. In fact, the exact values are not important because the behavior obtained will be the same as far as $n_{o}<n_{e}$.

We consider the propagation of a iniatially parallel light beam through the defect region in order to get a feeling for the bending of the rays by the defects. Figs. 5-9 show assorted lensing behavior from the defects studied. In particular, Fig. 5 shows a clear  lensing behavior for both $k=1$ defects, the $c=0$ defect acting as a diverging lens and the $c=\pi/2$ defect acting as a converging lens. The converging lens behavior is similar to the one presented by the cosmic string \cite{uzam} and by the irrotational hydrodynamical vortex \cite{visser}. The diverging lens behavior is similar to the one presented by a cosmic string with negative mass density \cite{uzam}. Like the vortex and differently from the cosmic string, the lensing effect here appears as consequence of the induced geometry, the effective geometry perceived by the light rays crossing the medium. The $k=-1$ and $k=\pm 1/2$ defects, being asymmetric, have different behavior depending on the incidence angle of the light, as shown in Figs. 6 - 9. 

The overall behavior observed in Figs. 5 - 9 is consistent with the the variation of the refractive index with the angle $\beta$ between the ray and the optical axis of the molecule. Notice that the optical length between two given points, which should be minimal, will be smaller for smaller $N_r$ (see Eqs. (\ref{fermat}) and (\ref{nr})). Therefore, the ray ``chooses'' to travel along the paths with smaller $N_r$, which in this case is parallel or antiparallel to the molecule (see Fig. 10). 

\begin{figure}[!h]
\begin{center}
\includegraphics[height=6cm]{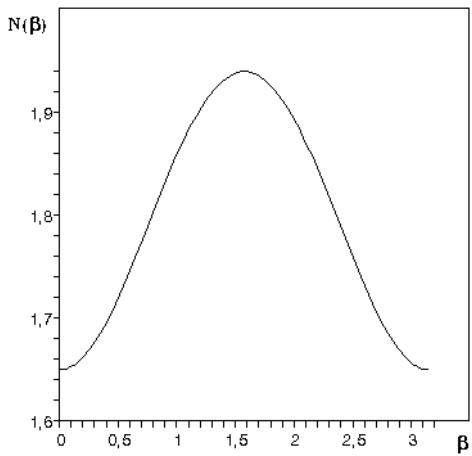} 
\caption{Refractive index $N_r$ as a function of the angle $\beta$ between the ray and the molecule's optical axis, from Eq. (\ref{nr}). $\beta$ is given in radians.}
\label{betafig} 
\end{center}
\end{figure}

\section{Concluding Remarks}
We study the propagation of light in a uniaxial liquid crystal for the specific molecular configurations corresponding to selected topological defects. We use a differential geometric approach, where the light rays are considered as geodesics in a generic orientation-dependent space. Naturally, the liquid crystal is in flat Euclidean space, only the bending of the extraordinary rays is interpreted as due to geometry. In other words,  we ask ourselves what is the geometry that gives the same paths as the ones followed by extraordinary light rays in nematics with topological defects. As a result we find a striking similarity with cosmic defects: the effective geometry associated to the hedgehog defect is that of the global monopole and that of the $k= 1$ disclinations is that of the cosmic string. The light paths near the defects are geodesics obtained from the geometry and were computed numerically showing an obvious lensing effect,  in agreement with the earlier result of Grandjean \cite{grand}. This similarity suggests that experiments to study optical properties of cosmic defects can be done with liquid crystals in the laboratory.

Analogue gravitational systems in condensed matter allow for  laboratory tests of some hypotheses otherwise of very difficult realization with present technology. Among these is gravitational lensing, an important research area in astronomy and astrophysics, which may benefit from the analogy shown in this article between cosmic defects and  defects in liquid crystals. A major difficulty in the research on gravitational lensing is the identification of the nature of a lens candidate. The possibility of simulating gravitational lensing behavior in the laboratory appears as a tool that might help in the characterization of known gravitational lenses like, for example, the CSL-1 (Capodimonte-Sternberg-Lens Candidate, No.1),  an extragalactic double source believed to be a cosmic string \cite{lens}. Although asymmetric cosmic strings are not yet known, a comparative study of the known multiple images formed by the many known gravitational lenses and the images formed by the $k=\pm \frac{1}{2}$ defects may indicate the presence of such objects.

\section{Appendix: Finslerian Approach}
In Finsler geometry \cite{finsler}, the line element depends not only on the position but also on the infinitesimal elements of direction $dx^i$, which can be written as $dx^i={\dot x}^i dt$, where $dt$ is a parameter along some curve. If we think of $t$ as time, ${\dot x}^i$ is the velocity of a particle moving along the curve which is therefore tangent to the curve. The arc length element is given by
\begin{equation}
ds =F(x,{\dot x})dt. \label{F}
\end{equation}
where $F(x,{\dot x})$ is the Finslerian function.
The geodesic equation in Finslerian geometry is
\begin{equation}
\frac{d^{2}x^i}{dt^2}+\Gamma^{i}_{jk}\frac{dx^j}{dt}\frac{dx^k}{dt}-\frac{d(log F)}{dt}\frac{dx^i}{dt}=0.\label{geo}
\end{equation}

In the Joets and Ribotta approach to geometric optics in anisotropic media \cite{joets} the
\begin{equation}
F(x,{\dot x})=N_r
\label{FN}
\end{equation}
identification is done.
Equation (\ref{geo}) can be simplified by noticing that $F$ will be constant ({\it i.e.} independent of $t$) and equal to $1$ if $t$ is taken to be the Finslerian arc length $s$ along the geodesic, which follows immediately by making $dt=ds$ in Eq. (\ref{F}). We then recover equation (\ref{georie}) of Riemannian geometry. The Christoffel symbols are computed in the same way as in Eq. (\ref{chris}) and the Finslerian metric tensor components,
\begin{equation}
g_{ij}=\frac{1}{2}\frac{\partial^{2}F^{2}}{\partial x^{i}\partial x^{j}}, \label{gtensor}
\end{equation}
turn out to be exactly the same ones computed with the help of Riemannian geometry for the cases studied. This justifies the use of Riemannian geometry for the calculations presented in the main body of this article.

\begin{acknowledgement} We thank CNPq, CAPES (PROCAD program) and PRONEX for financial support, H. Viglioni for the implementation of the Runge-Kutta routine and Profs. C. Furtado, T. Paschoal and J. Schaum  for important comments and suggestions.
\end{acknowledgement}

\end{document}